\documentclass[prd,nofootinbib,superscriptaddress]{revtex4}
\usepackage{amsmath,amssymb,graphicx}

\renewcommand{\vec}[1]{\boldsymbol{#1}}
\newcommand{\dif}{\mathrm{d}}

\begin{document}

\title{Exclusive photoproduction at the Tevatron and LHC within the dipole picture}
\author{L.~Motyka}
\affiliation{II Institute for Theoretical Physics, University of Hamburg, 22761 Hamburg, Germany}
\affiliation{Institute of Physics, Jagellonian University, 30-059 Krak\'ow, Poland}
\author{G.~Watt}
\affiliation{Department of Physics \& Astronomy, University College London, WC1E 6BT, UK}

\begin{abstract}
  We provide predictions for the rapidity distributions of exclusive photoproduced $J/\psi$ and $\Upsilon$ mesons, and $Z^0$ bosons, at the Tevatron and LHC.  We use the equivalent-photon approximation with the photon--proton cross sections given by the impact parameter dependent dipole saturation model, which has already been shown to give a good description of a wide variety of HERA data.  We derive the quark--antiquark light-cone wave functions of timelike neutral electroweak bosons. An essential difference is pointed out between the amplitude for timelike heavy boson photoproduction and the amplitude for deeply virtual Compton scattering.
\end{abstract}

\maketitle

\section{Introduction}
Exclusive diffractive vector meson production, $\gamma^{(*)}p\to V+p$, and deeply virtual Compton scattering (DVCS), $\gamma^*p\to \gamma+p$, have been extensively studied at HERA.  These processes provide a valuable probe of the generalised (or skewed) gluon density at small values of the proton's momentum fraction $x$~\cite{Ryskin:1992ui}.  Possible future measurements of the exclusive production of heavy vector mesons and $Z^0$ bosons in $p\bar p$ collisions at the Tevatron and $pp$ collisions at the LHC should extend and complement the existing measurements from HERA in terms of both the energy span and the precision.  In particular, by the measurement of exclusive $\Upsilon$ or $J/\psi$ production at the LHC through the detection of two rapidity gaps with the lepton pair from the meson decay identified in the central ATLAS or CMS detectors, one could probe the gluon density down to $x \sim 10^{-4}$ for $\Upsilon$ or $x\sim 10^{-5}$ for $J/\psi$.  Even smaller values of $x$ may be probed through more forward production at ALICE or LHCb.  Moreover, the event rates at the LHC are expected to be much higher than at HERA.  Exclusive $Z^0$ production at the LHC has two main merits: the process is perturbatively calculable with relatively small uncertainties and the experimental signature is very clean.  Therefore it provides an interesting probe of the interplay between strong and electroweak interactions in the diffractive channel.

\begin{figure}
  \centering
  \includegraphics[width=0.5\textwidth]{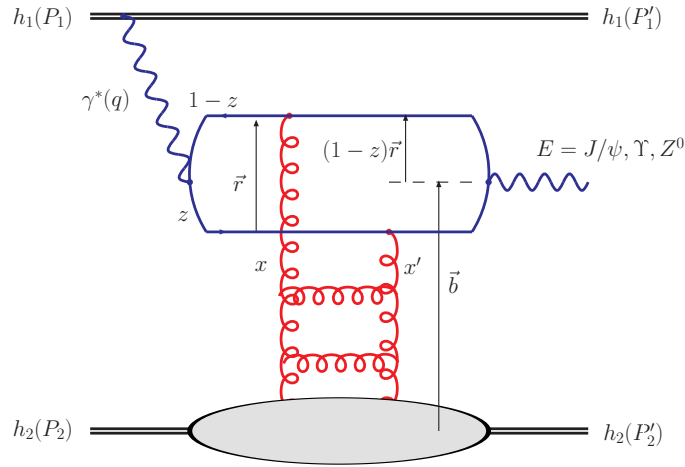}
  \caption{Exclusive photoproduction of vector mesons ($J/\psi,\Upsilon$) or $Z^0$ bosons in hadron--hadron interactions.  The diagram where the photon and the Pomeron are interchanged must also be included.  The variables in parentheses are the corresponding four-momenta.}
  \label{fig:diagram}
\end{figure}

The estimates of this paper are based on the equivalent-photon approximation combined with a dipole model used to compute the diffractive scattering of a quasireal photon on the proton.  In the presence of a hard scale, such as the heavy quark (or $Z^0$) mass, the impact parameter dependent dipole saturation (``b-Sat'') model \cite{Kowalski:2003hm,Kowalski:2006hc} incorporates the impact factor for the photon to heavy meson (or $Z^0$) transition in accordance with leading-order (LO) $k_\perp$-factorisation, together with LO DGLAP evolution of the gluon density.  The input gluon density was fitted to HERA data on the inclusive proton structure function $F_2$.  The model has been found to reproduce the main features of the HERA data for exclusive diffractive $J/\psi$, $\phi$ and $\rho$ production, and also for DVCS, as a function of the photon virtuality, $Q^2$, the photon--proton centre-of-mass energy, $W$, and the squared momentum transfer at the proton vertex, $t$ \cite{Kowalski:2003hm,Kowalski:2006hc,Watt:2007nr}.  Thus, in this paper we will use the b-Sat model to make predictions for the processes $h_1h_2\to h_1+E+h_2$ ($E=J/\psi,\Upsilon,Z^0$) at the Tevatron and LHC, where the reaction proceeds via photon--Pomeron fusion; see Fig.~\ref{fig:diagram}.  The quasireal photon can be emitted from either of the two incoming hadrons $h_i$.

In Sec.~\ref{sec:bsat} we recall the main formulae of the b-Sat model and in Sec.~\ref{sec:wavefunctions} we describe the wave functions used in the calculation.  We present the photon--proton cross sections in Sec.~\ref{sec:cross} and the hadron--hadron cross sections in Sec.~\ref{sec:rapidity}.  Finally, we compare our results with some other recent calculations in Sec.~\ref{sec:discussion} and conclude in Sec.~\ref{sec:conclusions}.  In the appendix we derive the light-cone wave functions for a timelike virtual photon or $Z^0$ boson.

\section{Impact parameter dependent dipole saturation model} \label{sec:bsat}
In this section we recall the main formulae involved in the application of the b-Sat model \cite{Kowalski:2003hm,Kowalski:2006hc} to the description of exclusive diffractive photoproduction.  The differential cross section for the exclusive process $\gamma p\to E+p$ is
\begin{equation}
  \frac{\dif\sigma^{\gamma p\rightarrow E+p}_{T}}{\dif t} = \frac{1}{16\pi}\left\lvert\mathcal{A}^{\gamma p\rightarrow E+p}_{T}\right\rvert^2,
  \label{eq:xvecm1}
\end{equation}
where the scattering amplitude
\begin{equation} \label{eq:exclamp}
  \mathcal{A}^{\gamma p\rightarrow E+p}_{T}(x,\vec{\Delta}) = \mathrm{i}\,\int\!\dif^2\vec{r}\int_0^1\!\frac{\dif{z}}{4\pi}\int\!\dif^2\vec{b}\;(\Psi_{E}^{*}\Psi_\gamma)_{T}\;\mathrm{e}^{-\mathrm{i}[\vec{b}-(1-z)\vec{r}]\cdot\vec{\Delta}}\;\frac{\dif\sigma_{q\bar q}}{\dif^2\vec{b}}\;\sqrt{(1+\beta^2)}.
\end{equation}
Here, $z$ is the fraction of the photon's light-cone momentum carried by the quark, $r=|\vec{r}|$ is the transverse size of the $q\bar{q}$ dipole, while $\vec{b}$ is the impact parameter, that is, $b$ is the transverse distance from the centre of the proton to the centre of mass of the $q\bar{q}$ dipole; see Fig.~\ref{fig:diagram}.  The transverse momentum lost by the outgoing proton, $\vec{\Delta}$, is the Fourier conjugate variable to the impact parameter $\vec{b}$, and $t=-\Delta^2$.  We assume $s$-channel helicity conservation in the $\gamma\to E$ transition.  We assume that the photon virtuality $Q^2\ll M_E^2$, then the contribution from longitudinally polarised photons may be neglected.  The factor $\exp[\mathrm{i}(1-z)\vec{r}\cdot\vec{\Delta}]$ in \eqref{eq:exclamp} originates from the nonforward wave functions \cite{Bartels:2003yj}.  The factor $\sqrt(1+\beta^2)$ in \eqref{eq:exclamp} is a correction to account for the imaginary part of the $S$-matrix element for the dipole--proton scattering, and is calculated using
\begin{equation} \label{eq:beta}
  \beta = \tan\left(\frac{\pi\lambda}{2}\right), \quad\text{with}\quad \lambda = \frac{\partial\ln\left(\dif\sigma_{q\bar q}/\dif^2\vec{b}\right)}{\partial\ln(1/x)}.
\end{equation}
The impact parameter dependent differential dipole cross section for the $q\bar{q}$ pair to scatter elastically off the proton is \cite{Kowalski:2006hc}
\begin{equation} \label{eq:dsigmad2b}
  \frac{\dif\sigma_{q\bar{q}}}{\dif^2\vec{b}} = 2\left[1-\exp\left(-\frac{\pi^2}{2N_c}r^2\alpha_S(\mu^2)\,R_g\,xg(x,\mu^2)\,T(b)\right)\right].
\end{equation}
Here, the scale $\mu^2$ is related to the dipole size $r$ by $\mu^2=4/r^2+\mu_0^2$.  The gluon density, $xg(x,\mu^2)$, is evolved from a scale $\mu_0^2$ up to $\mu^2$ using LO DGLAP evolution without quarks.  The factor $R_g$ in \eqref{eq:dsigmad2b} accounts for the skewedness effect, that is, $x\ne x^\prime$ in Fig.~\ref{fig:diagram}, and is calculated using \cite{Shuvaev:1999ce}
\begin{equation} \label{eq:Rg}
  R_g(\lambda_s) = \frac{2^{2\lambda_s+3}}{\sqrt{\pi}}\frac{\Gamma(\lambda_s+5/2)}{\Gamma(\lambda_s+4)}, \quad\text{with}\quad \lambda_s = \frac{\partial\ln\left[xg(x,\mu^2)\right]}{\partial\ln(1/x)}.
\end{equation}
The definitions of $\lambda$ and $\lambda_s$ given by \eqref{eq:beta} and \eqref{eq:Rg}, respectively, are formally equivalent in the colour transparency limit ($r\to 0$), and are numerically very similar for the observables computed in this paper.\footnote{Note that the definition of $\lambda$ given by \eqref{eq:beta} differs from that used in Ref.~\cite{Kowalski:2006hc}, where $\lambda$ was calculated as the logarithmic derivative of the $\gamma p$ amplitude rather than the dipole cross section.  The definition of $\lambda$ in \eqref{eq:beta} is more convenient for the case of $Z^0$ photoproduction where there are both real and imaginary parts to the $\gamma\to Z^0$ impact factor.}
The initial gluon density at the scale $\mu_0^2$ is taken in the form
\begin{equation} \label{eq:inputgluon}
  xg(x,\mu_0^2) = A_g\,x^{-\lambda_g}\,(1-x)^{5.6}.
\end{equation}
The values of the parameters $\mu_0^2$, $A_g$, and $\lambda_g$ were determined from a fit to HERA $F_2$ data \cite{Kowalski:2006hc}.  The dipole cross section is evaluated at $x=M_E^2/W^2$ for $E=J/\psi,\Upsilon,Z^0$.  The proton shape function $T(b)$ takes a Gaussian form, that is,
\begin{equation} \label{eq:GaussianTb}
  T(b) = \frac{1}{2\pi B_G}\mathrm{e}^{-\frac{b^2}{2B_G}},
\end{equation}
where $B_G = 4$ GeV$^{-2}$ is determined by the comparison to data for the $t$ dependence of exclusive $J/\psi$ photoproduction at HERA \cite{Kowalski:2006hc}.  Note that although the b-Sat model incorporates saturation effects via the eikonalisation of the gluon density in \eqref{eq:dsigmad2b}, these saturation effects are expected to be only moderate for $J/\psi$ production and negligible for $\Upsilon$ and $Z^0$ production, since the scattering amplitudes are dominated by increasingly small dipole sizes with increasing mass of the produced particle.

\section{Photon, meson and $Z^0$ wave functions} \label{sec:wavefunctions}

The forward overlap function between the transversely polarised photon and vector meson wave functions in \eqref{eq:exclamp}, $(\Psi_V^*\Psi_\gamma)_{T}$, is given in Ref.~\cite{Kowalski:2006hc}.  We use the ``boosted Gaussian'' vector meson wave functions \cite{Nemchik:1994fp,Nemchik:1996cw,Forshaw:2003ki}, which were found to give the best description of HERA data \cite{Kowalski:2006hc}.  The parameters for the $J/\psi$ and $\Upsilon$ wave functions are given in Table~\ref{tab:bGparams}.
\begin{table}
  \centering
  \begin{tabular}{cccc|cccc}
    \hline\hline
    Meson & $M_V$/GeV & $f_V$/GeV & $m_f$/GeV & $\mathcal{N}_T$& $\mathcal{N}_L$& $\mathcal{R}^2$/GeV$^{-2}$ & $f_{V,T}$/GeV \\ \hline
    $J/\psi$ & 3.097 & 0.274 & 1.4 & 0.578  & 0.575  & 2.3 & 0.307 \\\hline
    $\Upsilon(1S)$ & 9.460 & 0.236 & 4.5 & 0.469 & 0.469 & 0.55 & 0.252 \\
    $\Upsilon(1S)$ & 9.460 & 0.236 & 4.2 & 0.481 & 0.480 & 0.57 & 0.238 \\
    \hline\hline
  \end{tabular}
  \caption{Parameters of the ``boosted Gaussian'' vector meson wave functions; see Ref.~\cite{Kowalski:2006hc} for their definitions.  For $\Upsilon$ the parameters are given for two different values of the bottom quark mass $m_b$.}
  \label{tab:bGparams}
\end{table}

For $Z^0$ production, the amplitude \eqref{eq:exclamp} involves a sum over quark flavours $f=u,d,s,c,b$.  The wave functions for an incoming $Z^0$ with spacelike virtuality $q^2=-Q^2<0$ are known from the application of the colour dipole picture to charged-current deep-inelastic scattering \cite{Fiore:2005yi,Fiore:2005bp}.  The wave functions for an outgoing $Z^0$ with timelike $q^2 = M_Z^2 >0$ are derived in the appendix.  The transversely polarised overlap function between the (quasireal) photon wave function and the $Z^0$ wave function for quark flavour $f$ is given by
\begin{equation}
  (\Psi_{Z^0}^*\Psi_{\gamma})_{T}^f = 
\frac{2 N_c\,\alpha_{\rm em}}{\pi} \,
\frac{e_f\,g_v ^f}{\sin 2\theta_W} \,
\left\{\left[z^2+(1-z)^2\right]m_f K_1(m_f r) 
\tilde\varepsilon_Z K_1(\tilde\varepsilon_Z r)
\,+\, m_f^2 K_0(m_f r) K_0(\tilde\varepsilon_Z r) \right\}.
  \label{eq:overlap_Zgamma}
\end{equation}
Here, the vector couplings are $g_v ^{u,c}=1/2-4/3\,\sin^2\theta_W$ and $g_v ^{d,s,b}=-1/2+2/3\,\sin^2\theta_W$ where $\theta_W$ is the Weinberg angle, and
\begin{equation} \label{eq:epsilonZ}
  \tilde\varepsilon_Z = \begin{cases}\sqrt{m_f^2 - M_Z^2\,z(1-z)} & :\quad m_f^2 - M_Z^2\,z(1-z) > 0\\
    -{\rm i}\sqrt{M_Z^2\,z(1-z)-m_f^2} & :\quad M_Z^2\,z(1-z) -m_f^2 > 0
  \end{cases}.
\end{equation}
The default quark masses are taken to be $m_{u,d,s} = 0.14$ GeV, $m_c = 1.4$ GeV and $m_b = 4.5$ GeV.  We use a fixed value of $\alpha_{\rm em} = 1/137$ for the numerical results presented in this paper, although it may be more appropriate to use a running $\alpha_{\rm em}(M_Z^2) \simeq 1/128$, in which case the $Z^0$ cross sections would increase by 15\%.

The quasireal photon--proton cross section for $Z^0$ production at a centre-of-mass energy $W$ is therefore identical to the cross section for timelike Compton scattering, $\gamma p\to\gamma^* p$, at a produced photon virtuality $q^2=M_Z^2$, apart from the replacement of the coupling in the amplitude:
\begin{equation}
  e\,e_f \quad \longrightarrow \quad \frac{e\,g_v^f}{\sin 2\theta_W}.
\end{equation}
Note that the timelike Compton scattering process, $\gamma p\to\gamma^* p$, has so far only been studied at LO in the collinear factorisation framework in terms of the generalised quark distribution \cite{Berger:2001xd}.  The timelike Compton scattering process at the LHC will be sensitive to the generalised gluon distribution \cite{Pire:2008pm}, and this process is calculable within the dipole picture using the wave functions for a timelike virtual photon given in the appendix.

When evaluating the modified Bessel functions of an imaginary argument it is convenient to use the following relations of Bessel functions, valid for a real variable $x>0$:
\begin{equation}
K_0(-ix) \,=\,  -\frac{\pi}{2}\, [\, Y_0(x)\, - \, iJ_0(x)\,], \quad
K_1(-ix) \,=\,  -\frac{\pi}{2}\, [\, J_1(x)\, + \, iY_1(x)\,].
\end{equation}
Note that the wave function of the timelike vector boson essentially differs from the wave function of the virtual spacelike boson.  The origin of this difference is kinematic: this point is explained in detail in the appendix.  As a consequence, the $Z^0$ photoproduction amplitude does not equal the electroweak DVCS amplitude at $Q^2=M_Z^2$. In particular, one sees that the overlap function, given by (\ref{eq:overlap_Zgamma}), picks up an imaginary part related to the contribution of an on-shell quark--antiquark pair at the $Z^0$ vertex.

\begin{figure}
  \centering
  \includegraphics[width=0.33\textwidth]{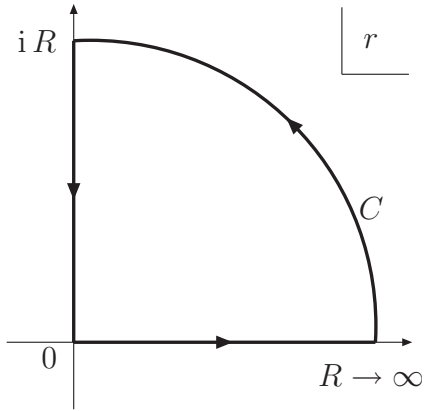}%
   \caption{Choice of integration contour $C$ to evaluate the amplitude for $\gamma p\to Z^0+p$ when $|\tilde\varepsilon_Z|>m_f$.}
  \label{fig:contour}
\end{figure}

\begin{figure}
  \begin{minipage}{\textwidth}
    \centering
    $\int\limits_0^\infty\!\dif{r}$
    \begin{minipage}{0.4\textwidth}
      \includegraphics[width=\textwidth,clip]{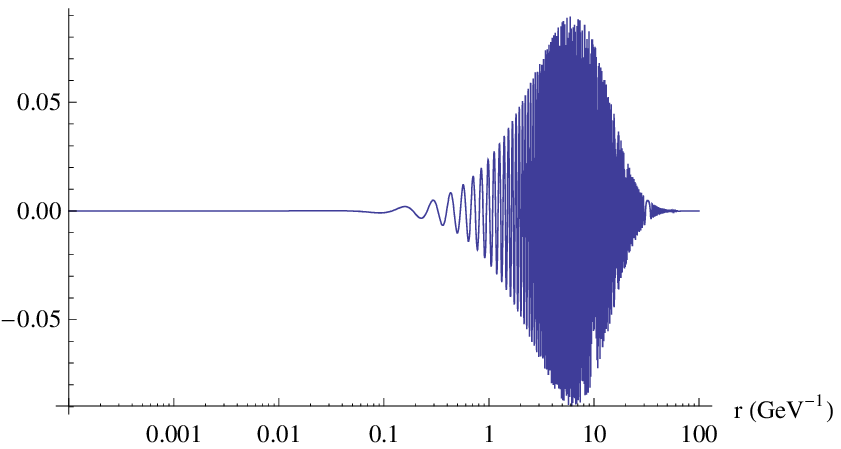}
    \end{minipage} = $\int\limits_0^\infty\!\dif{r}$
    \begin{minipage}{0.4\textwidth}
      \includegraphics[width=\textwidth,clip]{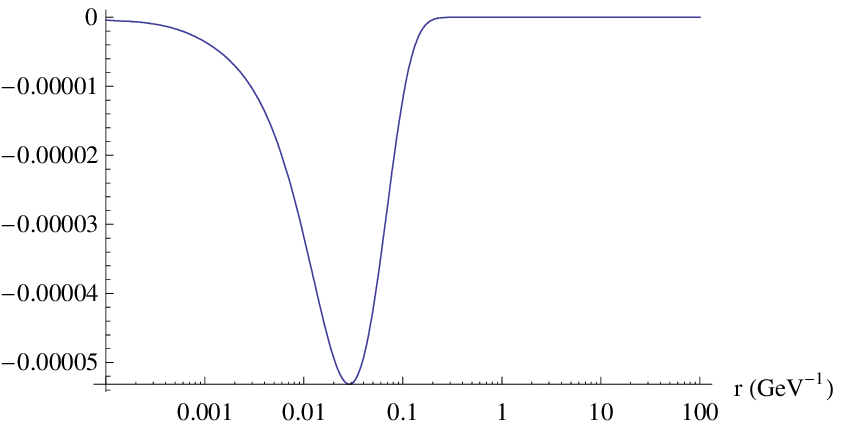}
    \end{minipage}
  \end{minipage}
  \caption{Application of contour integration \eqref{eq:contour} to evaluate the imaginary part of the down quark contribution to the amplitude for $\gamma p\to Z^0+p$ for a fixed $z=0.5$ with $\Delta=0$ and $x=M_Z/\sqrt{s}$ corresponding to central production at the LHC.}
  \label{fig:integration}
\end{figure}

On substituting the overlap function \eqref{eq:overlap_Zgamma} into the amplitude \eqref{eq:exclamp}, one finds that the integrand is wildly oscillatory as a function of $r$ if $|\tilde\varepsilon_Z|\gg m_f$, meaning that direct numerical integration over $r$ is difficult.  This problem can be solved by taking the analytic continuation to complex $r$.  By observing that the integrand is much better behaved under the replacement $r\to\mathrm{i}r$ one can choose the integration contour shown in Fig.~\ref{fig:contour}.
If there are no poles inside the contour $C$, then application of the residue theorem gives
\begin{equation}
  0 = \oint_C\!\dif{r}\,f(r) = \int_0^R\!\dif{r}\,f(r) + \int_0^{\pi/2}\!\dif{\theta}\,{\rm i}R{\rm e}^{{\rm i}\theta}f(R{\rm e}^{{\rm i}\theta}) + \int_{{\rm i}R}^0\!\dif{({\rm i}r)}\,f({\rm i}r).
\end{equation}
Taking the limit $R\to\infty$ gives the result that
\begin{equation} \label{eq:contour}
  \int_0^\infty\!\dif{r}\;f(r) = \int_0^\infty\!\dif{r}\;{\rm i}f({\rm i}r).
\end{equation}
We use this technique to evaluate the integral over $r$ for the case of $|\tilde\varepsilon_Z|>m_f$.  (If $|\tilde\varepsilon_Z|<m_f$, the integral over $r$ can be done in the usual way.  Note that there is an integrable singularity at $\tilde\varepsilon_Z=0$.)  This technique requires an analytic form of the dipole cross section as a function of $r$.  This can be obtained by fitting the quantity
\begin{equation}
  \frac{1}{r^2}\int\!\dif^2\vec{b}\;\mathrm{e}^{-\mathrm{i}\vec{b}\cdot\vec{\Delta}}\;\frac{\dif\sigma_{q\bar q}}{\dif^2\vec{b}}\;\sqrt{(1+\beta^2)},
\end{equation}
for fixed values of $x$ and $\Delta$, to a polynomial of degree 15 in $\log_{10}(r)$ for $r\in[10^{-4},10^2]$ GeV$^{-1}$.  (Too few terms in the polynomial will mean that the form of the dipole cross section is not well reproduced, while too many terms will mean that numerical rounding errors become sizeable when $r\to{\rm i}r$ is taken due to partial cancellation between the different terms.)  The application of equation \eqref{eq:contour} for a fixed $z=0.5$ is illustrated in Fig.~\ref{fig:integration}.  It is seen that the integrand is much better behaved under $r\to{\rm i}r$ such that numerical integration over $r$ is then straightforward.

\section{Photon--proton cross sections} \label{sec:cross}

\begin{figure}
  \centering
  \includegraphics[width=0.89\textwidth]{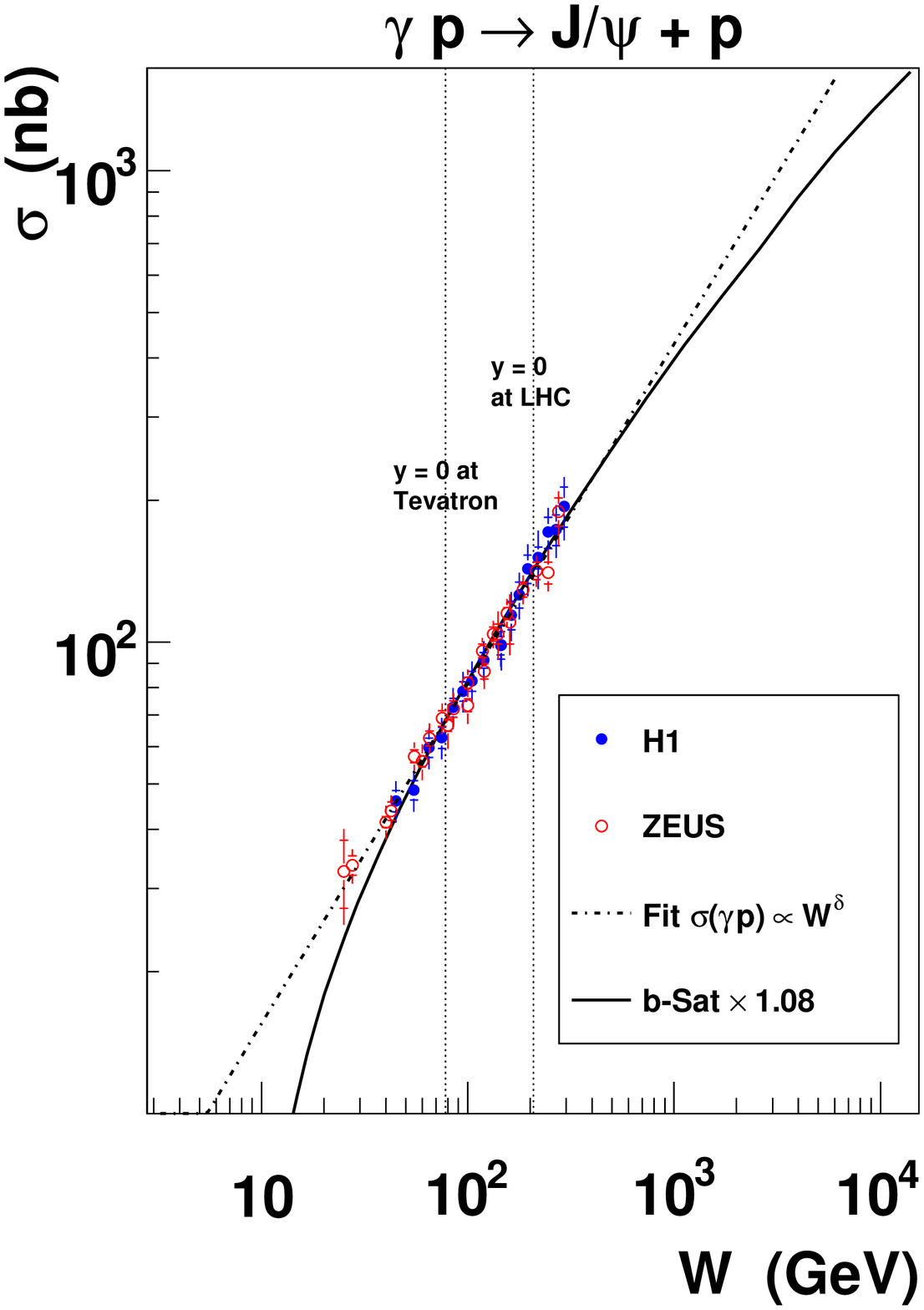}
  \caption{$W$ dependence of the $\gamma p$ cross section for exclusive photoproduction of $J/\psi$ mesons.  The ``b-Sat'' model predictions are rescaled by a factor 1.08 to give optimum agreement with the HERA data \cite{Chekanov:2002xi,Aktas:2005xu}.  Also shown is a direct fit to the HERA data of the form $\sigma(\gamma p) \propto W^\delta$.  The values of $W$ corresponding to central production at the Tevatron and LHC are indicated.}
  \label{fig:sigmaw_jpsi}
\end{figure}
\begin{figure}
  \centering
  \includegraphics[width=0.89\textwidth]{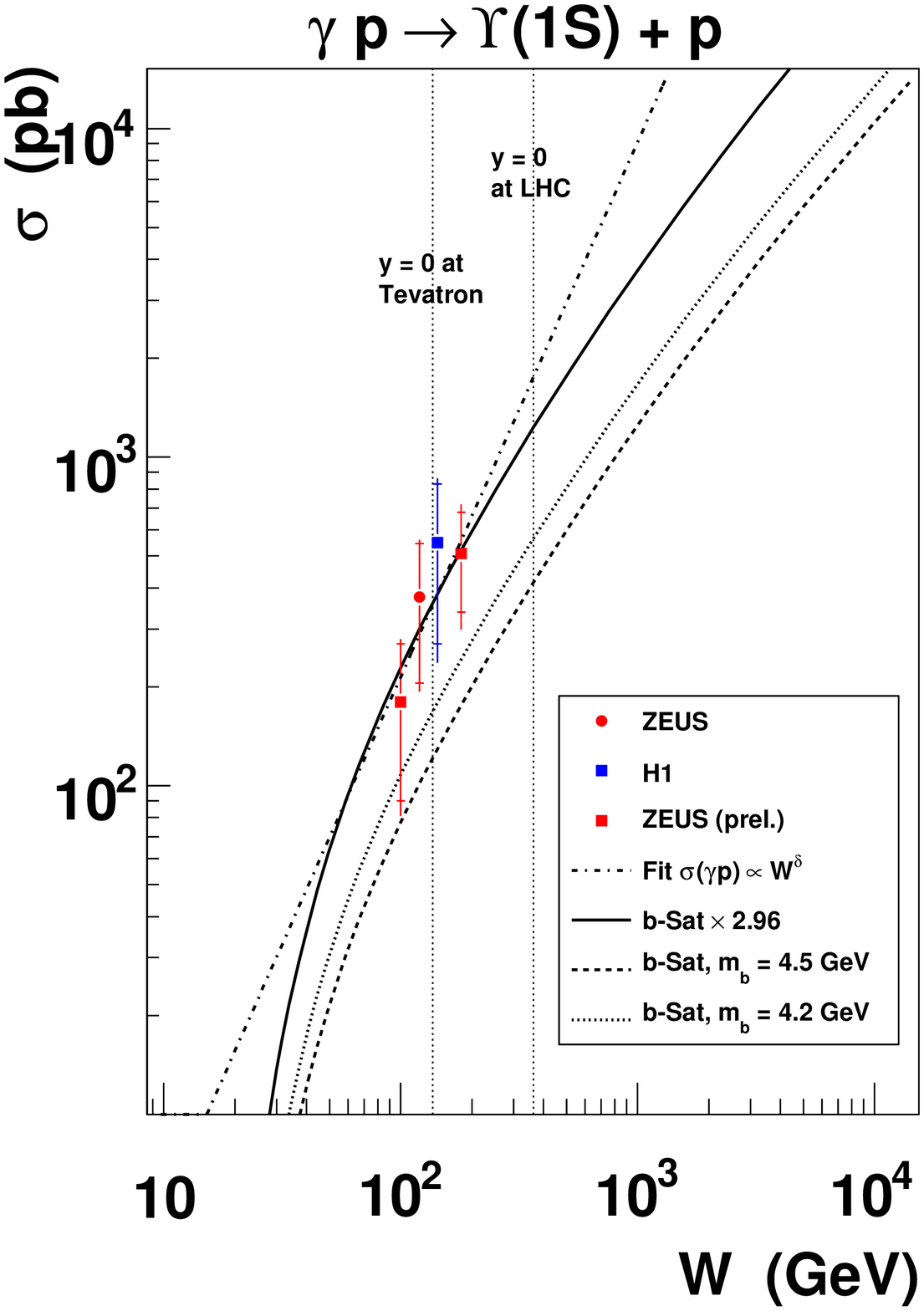}
  \caption{$W$ dependence of the $\gamma p$ cross section for exclusive photoproduction of $\Upsilon$ mesons.  The ``b-Sat'' model predictions with $m_b = 4.5$ GeV are rescaled by a factor 2.96 to give optimum agreement with the HERA data \cite{Breitweg:1998ki,Adloff:2000vm,ZEUSupsilon}.  Also shown is a direct fit to the HERA data of the form $\sigma(\gamma p) \propto W^\delta$.  The values of $W$ corresponding to central production at the Tevatron and LHC are indicated.}
  \label{fig:sigmaw_upsilon}
\end{figure}
\begin{figure}
  \centering
  \includegraphics[width=0.89\textwidth]{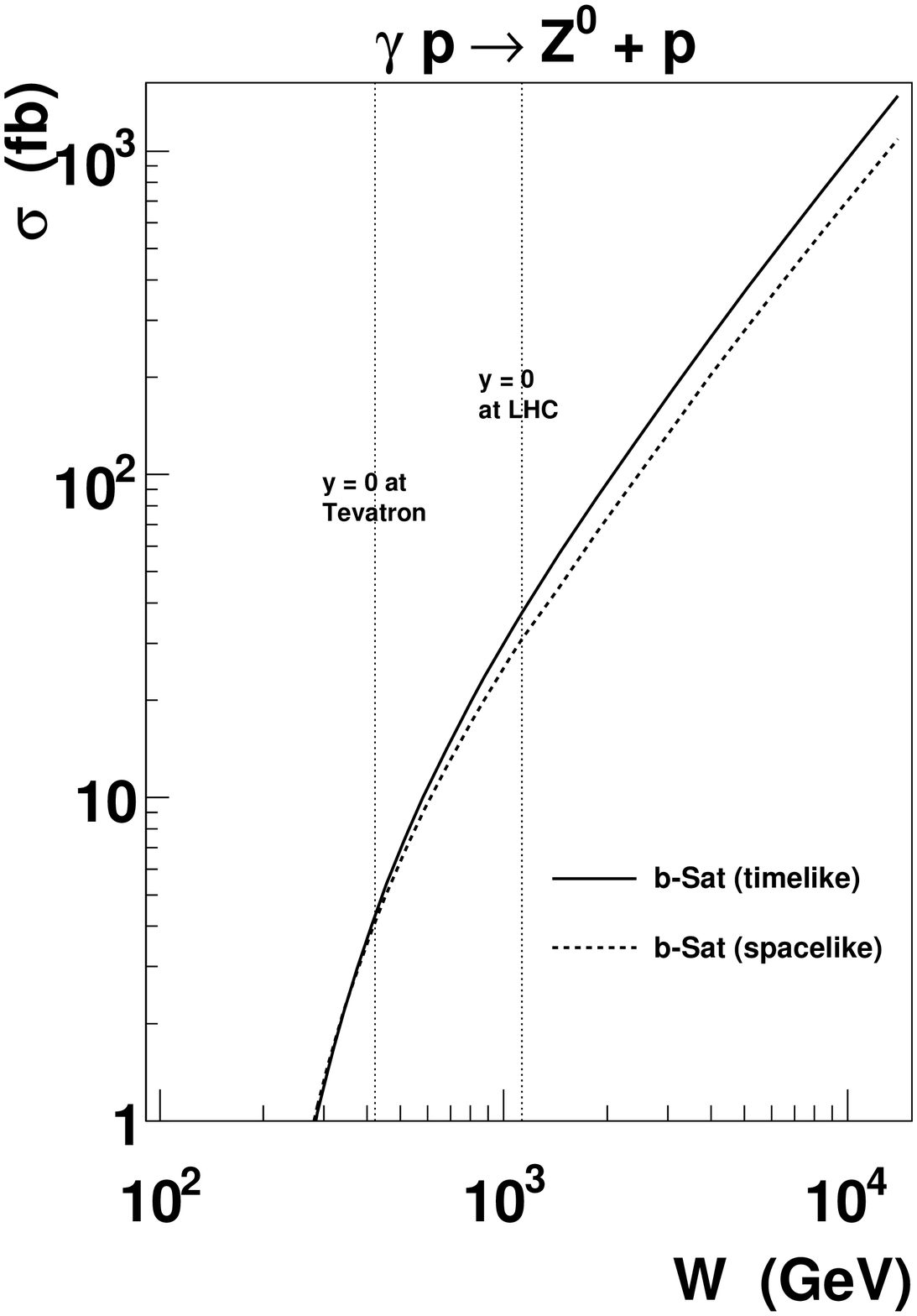}%
  \caption{$W$ dependence of the $\gamma p$ cross section for exclusive photoproduction of $Z^0$ bosons in the (correct) timelike case and the (incorrect) spacelike case.  The values of $W$ corresponding to central production at the Tevatron and LHC are indicated.}
  \label{fig:sigmaw_z}
\end{figure}
In Figs.~\ref{fig:sigmaw_jpsi}, \ref{fig:sigmaw_upsilon} and \ref{fig:sigmaw_z} we show the $\gamma p$ cross sections calculated using \eqref{eq:xvecm1}, integrated over $|t|<1$ GeV$^2$, for $W$ up to the maximum value of $\sqrt{s} = 14$ TeV theoretically accessible at the LHC.  We indicate the values of the photon--proton centre-of-mass energy $W$ probed by central production at the Tevatron and LHC.  The predictions are compared to the available HERA data for exclusive $J/\psi$ \cite{Chekanov:2002xi,Aktas:2005xu} and $\Upsilon$ \cite{Breitweg:1998ki,Adloff:2000vm,ZEUSupsilon} photoproduction.  For $J/\psi$ production there is good agreement of the predictions with the data, as already observed in Ref.~\cite{Kowalski:2006hc}, but to obtain optimum agreement with the HERA data for the purposes of providing predictions for the Tevatron and LHC we scale the predictions for the $\gamma p$ cross section by an overall normalisation factor of 1.08.

For $\Upsilon$ production, however, the predictions lie roughly a factor 2--3 below the data points for our default choice of bottom quark mass $m_b = 4.5$ GeV; see Fig.~\ref{fig:sigmaw_upsilon}.  The corrections for skewedness and the real part of the amplitude are large, at about a factor 2--3 for $\Upsilon$, but these are already included in the ``b-Sat'' predictions.  The $\Upsilon$ predictions are found to be sensitive to the assumed value of the bottom quark mass, as seen by the alternative predictions shown in Fig.~\ref{fig:sigmaw_upsilon} with $m_b = 4.2$~GeV corresponding to the running $\overline{\rm MS}$ mass, $m_b(m_b)$ \cite{Yao:2006px}.  Even with this low mass value, however, the theory curves lie significantly below the data.  Such discrepancy of theory predictions and the HERA $\Upsilon$ data is typically found in models that use LO accuracy and an explicit meson wave function, constrained by the leptonic decay width; see, for example, Ref.~\cite{Rybarska:2008pk}.  The successful Martin--Ryskin--Teubner (MRT) description of HERA $\Upsilon$ data obtained in~Ref.~\cite{Martin:1999rn}, although similar in spirit, is distinct from our approach: the $b\bar{b}\to\Upsilon$ transition is modelled by MRT using either the meson distribution amplitude or the parton--hadron duality hypothesis; see also Ref.~\cite{Martin:2007sb}.  From the discrepancy between the MRT results and those from our approach, one may conclude that the $\Upsilon$ cross section is strongly sensitive to the choice of the $\Upsilon$ wave function.  Indeed, it was demonstrated \cite{Rybarska:2008pk} that alternative wave functions to the ``boosted Gaussian'' used here give a large spread in the predictions.  In addition, in our analysis we do not account for the higher-order QCD corrections to the impact factor, which can be quite large \cite{Ivanov:2004vd,Ivanov:2005gn}.  All these effects, however, are expected to mainly alter the overall normalisation, but not the $W$ dependence which is given by the gluon density or dipole cross section.  Therefore, rather than confront these issues, we simply rescale the b-Sat model predictions with $m_b = 4.5$~GeV by a factor 2.96 to provide optimum agreement with the available HERA data.  The rescaled prediction is also shown in Fig.~\ref{fig:sigmaw_upsilon}.

As well as the b-Sat model predictions, we will also show the results of simply fitting the HERA $J/\psi$ and $\Upsilon$ data to the powerlike form $\sigma(\gamma p\to V+p) \propto W^{\delta}$, which gives $\sigma(\gamma p\to J/\psi+p) = (3.0\,\mathrm{nb})(W/W_0)^{0.72}$ and $\sigma(\gamma p\to \Upsilon+p) = (0.12\,\mathrm{pb})(W/W_0)^{1.6}$ with $W_0 = 1$ GeV, shown by the dot-dashed curves in Figs.~\ref{fig:sigmaw_jpsi} and \ref{fig:sigmaw_upsilon}.  It is clear that any extrapolation based on the four imprecise $\Upsilon$ data points will have a large uncertainty.  In particular, note that the value of $W$ probed at central rapidity at the LHC is outside the HERA kinematic range.  Note also that these fits are to more precise HERA data than were available in Ref.~\cite{Klein:2003vd}.

To investigate the numerical impact of the timelike kinematics on the cross section for $Z^0$ we also calculated the cross section with (incorrect) spacelike kinematics, that is, with $M_Z^2\to -M_Z^2$ in \eqref{eq:epsilonZ}, then the amplitude is the same as that for DVCS at a scale $Q^2=M_Z^2$ apart from the different coupling.  The magnitude of the imaginary part of the amplitude is very similar in both the timelike and spacelike cases: in the timelike case for central production it is 0.5\% smaller at the Tevatron and 2.8\% larger at the LHC compared to the spacelike case.  However, there is also a significant \emph{real} part of the amplitude in the timelike case which is 24\% (Tevatron) and 38\% (LHC) of the imaginary part.  The cross sections at $y=0$ are therefore enhanced by 5\% at the Tevatron and 21\% at the LHC in the timelike case compared to the spacelike case.

\begin{table}
  \centering
  \begin{tabular}{c|c|c|c|c}
    \hline\hline
    $J/\psi$, $y=0$ & $W_0$ (GeV) & $\sigma^{\gamma p}(W_0)$ (nb) & $\delta$ & $B_D$ (GeV$^{-2}$) \\ \hline
    Tevatron & $78$ & $68$ & $0.80$ & $4.66$ \\
    LHC & $208$ & $142$ & $0.71$ & $4.72$ \\
    \hline\hline\multicolumn{5}{c}{}\\\hline\hline
    $\Upsilon(1S)$, $y=0$ & $W_0$ (GeV) & $\sigma^{\gamma p}(W_0)$ (pb) & $\delta$ & $B_D$ (GeV$^{-2}$) \\ \hline
    Tevatron & $136$ & $360$ & $1.39$ & $4.12$ \\
    LHC & $364$ & $1233$ & $1.16$ & $4.15$ \\
    \hline\hline\multicolumn{5}{c}{}\\\hline\hline
    $Z^0$, $y=0$ & $W_0$ (GeV) & $\sigma^{\gamma p}(W_0)$ (fb) & $\delta$ & $B_D$ (GeV$^{-2}$) \\ \hline
    Tevatron & $423$ & $4.2$ & $3.03$ & $4.25$ \\
    LHC & $1130$ & $37$ & $1.73$ & $4.17$ \\
    \hline\hline
  \end{tabular}
  \caption{Values of the ``b-Sat'' model predictions for the $\gamma p$ cross section corresponding to central production at the Tevatron and LHC.  The $J/\psi$ and $\Upsilon$ predictions have been scaled by factors 1.08 and 2.96, respectively, in order to give the best agreement with the existing HERA data; see Figs.~\ref{fig:sigmaw_jpsi} and \ref{fig:sigmaw_upsilon}.  A reasonable approximation of the $W$ and $t$ dependence in the vicinity of $W_0$ may be obtained from $\dif\sigma^{\gamma p}/\dif t = \sigma^{\gamma p}(W_0)\,(W/W_0)^\delta\,B_D\,\exp(-B_D|t|)$.}
  \label{tab:sigmaw}
\end{table}

For the convenience of possible future studies, we provide a simple parameterisation of the b-Sat model predictions for the $\gamma p$ cross sections shown in Figs.~\ref{fig:sigmaw_jpsi}, \ref{fig:sigmaw_upsilon} and \ref{fig:sigmaw_z} corresponding to central production at the Tevatron and LHC.  In Table~\ref{tab:sigmaw} we give the values of $W=W_0$ at $y=0$, the power $\delta = \partial\ln\sigma^{\gamma p}/\partial\ln W|_{W=W_0}$ characterising the $W$ dependence, and the $t$-slope parameter $B_D$ obtained by fitting $\dif\sigma^{\gamma p}/\dif t\propto\,\exp(-B_D|t|)$ for $|t|<1$ GeV$^2$.  A reasonable approximation of the $W$ and $t$ dependence of the $\gamma p$ cross section in the vicinity of $W_0$ may therefore be obtained from
\begin{equation} \label{eq:param}
  \frac{\dif\sigma^{\gamma p}}{\dif t} = \sigma^{\gamma p}(W_0)\,\left(\frac{W}{W_0}\right)^\delta\,B_D\,\exp(-B_D|t|).
\end{equation}
A comment on the uncertainty of this parameterisation is in order.  For $J/\psi$ production, the accuracy of the HERA data and the agreement between these data and the b-Sat model suggests that the uncertainty of the theory predictions for the $\gamma p \to J/\psi+p$ cross section is ${\cal O}(10\%)$.  At this level of accuracy, one should explicitly impose the rescattering correction of $\sim$ 0.7--0.9 when applying the parameterisation \eqref{eq:param} to hadron--hadron collisions; see the later discussion in Sec.~\ref{sec:discussion}.  For $\Upsilon$ production, the large experimental errors on the HERA data points and the spread between the various theory predictions suggests that the normalisation uncertainty factor is about 2--3; however, the energy dependence is expected to be accurately predicted.  The estimate for $Z^0$ production is theoretically cleanest: here we expect the relative uncertainty on the predictions to come mostly from higher-order QCD corrections, so to be ${\cal O}(\alpha_S)$, bearing in mind, however, that the numerical prefactor of $\alpha_S$ in the next-to-leading-order correction may easily be greater than one.

\section{Rapidity distributions at the Tevatron and LHC} \label{sec:rapidity}

To obtain the hadron--hadron cross sections from the photon--proton cross sections, we need to multiply by the photon flux $\dif n/\dif k$ and integrate over the photon energy $k$ \cite{Klein:2003vd}:
\begin{equation} \label{eq:xsec}
  \sigma(h_1h_2\to h_1+E+h_2) = 2\int_0^\infty\!\dif{k}\;\frac{\dif n}{\dif k}\;\sigma(\gamma p\to E+p).
\end{equation}
The initial factor of 2 in \eqref{eq:xsec} accounts for the interchange of the photon emitter and the target.  We neglect the absorptive corrections due to spectator interactions between the two hadrons and will comment on the effects of these in Sec.~\ref{sec:discussion}. The possible interference between photon--Pomeron and Pomeron--photon fusion has a large effect only for very small meson transverse momenta~\cite{Klein:1999gv} and may be safely neglected in our analysis; see also~Ref.~\cite{Schafer:2007mm}.

The four-momentum of the exchanged photon in Fig.~\ref{fig:diagram} is $q=(k,\vec{q}_\perp,k/\beta_L)$, where $k$ and $\vec{q}_\perp$ are the energy and transverse momentum of the quasireal photon in a given frame, where the projectile moves with velocity $\beta_L$ \cite{Baur:2001jj}.  Therefore, the photon virtuality is $q^2 = -Q^2 = -k^2/(\gamma_L^2\,\beta_L^2)-q_\perp^2$, where $\gamma_L=(1-\beta_L^2)^{-1/2}=\sqrt{s}/(2m_p)$ is the Lorentz factor of a single beam.  The photon energy spectrum is given by a modified equivalent-photon (Weizs\"acker--Williams) approximation \cite{Klein:2003vd,Drees:1988pp}:
\begin{equation} \label{eq:photonflux}
  \frac{\dif n}{\dif k} = \frac{\alpha_{\rm em}}{2\pi k}\left[1+\left(1-\frac{2k}{\sqrt{s}}\right)^2\right]\left(\ln A-\frac{11}{6}+\frac{3}{A}-\frac{3}{2A^2}+\frac{1}{3A^2}\right),
\end{equation}
where $A=1+(0.71\,{\rm GeV}^2)/Q_{\rm min}^2$ and $Q_{\rm min}^2 \simeq k^2/\gamma_L^2$.  This result \eqref{eq:photonflux} has been obtained by integrating over the product of the photon propagator, $1/Q^2$, and the squared electromagnetic form factor of the proton, $F^2(Q^2) = (1+Q^2/(0.71\,{\rm GeV}^2))^{-4}$ \cite{Drees:1988pp}.  We neglect the virtuality $Q^2$ of the quasireal photon wherever possible in the calculation of the $\gamma p$ subprocess.  The square of the $\gamma p$ centre-of-mass energy, $W^2\simeq 2k\sqrt{s}$, where $\sqrt{s}$ is the hadron--hadron centre-of-mass energy.  The produced state with mass $M_E$ has rapidity $y\simeq \ln(2k/M_E)$, so \eqref{eq:xsec} can be rewritten as \cite{Klein:2003vd}
\begin{equation} \label{eq:rapdist}
  \frac{\dif \sigma}{\dif y}(h_1h_2\to h_1+E+h_2) = k\frac{\dif n}{\dif k}\sigma(\gamma p\to E+p)\quad+\quad(y\to -y),
\end{equation}
where the photon energy $k\simeq (M_E/2)\exp(y)$.  Neglecting interference, the contribution from the interchange of the photon emitter and the target can be obtained by replacing $y\to -y$.

In Table \ref{tab:results}, we give the hadron--hadron cross sections at central rapidity for the Tevatron and LHC, the total cross sections integrated over rapidity, and the total event rates including the appropriate leptonic branching ratios and assuming the Tevatron and LHC design luminosities.  The cross sections for $Z^0$ production are comparable with the cross section predictions for exclusive diffractive Higgs ($M_H = 120$ GeV) production of $0.2$ fb (Tevatron) and $3$ fb (LHC) \cite{Khoze:2001xm}.
\begin{table}
  \centering
  \begin{tabular}{c|c|c|c}
    \hline\hline
    $J/\psi$ & $\dif\sigma/\dif y|_{y=0}$ (nb) & $\sigma$ (nb) & Event rate (s$^{-1}$) \\ \hline
    Tevatron & $3.4$ & $28$ & $0.33$ \\
    LHC & $9.8$ & $120$ & $71$ \\
    \hline\hline\multicolumn{4}{c}{}\\\hline\hline
    $\Upsilon(1S)$ & $\dif\sigma/\dif y|_{y=0}$ (pb) & $\sigma$ (pb) & Event rate (hr$^{-1}$) \\ \hline
    Tevatron & $14$ & $115$ & $2.0$ \\
    LHC & $72$ & $1060$ & $946$ \\
    \hline\hline\multicolumn{4}{c}{}\\\hline\hline
    $Z^0$ & $\dif\sigma/\dif y|_{y=0}$ (fb) & $\sigma$ (fb) & Event rate (yr$^{-1}$) \\ \hline
    Tevatron & $0.077$ & $0.30$ & $0.065$ \\
    LHC & $1.4$ & $13$ & $135$ \\
    \hline\hline
  \end{tabular}
  \caption{``b-Sat'' model predictions for $J/\psi$, $\Upsilon$ and $Z^0$ photoproduction at the Tevatron Run II ($\sqrt{s}=1.96$ TeV) and the LHC ($\sqrt{s}=14$ TeV) using the equivalent-photon approximation.  The $J/\psi$ and $\Upsilon$ predictions have been scaled by factors 1.08 and 2.96, respectively, in order to give the best agreement with the existing HERA data; see Figs.~\ref{fig:sigmaw_jpsi} and \ref{fig:sigmaw_upsilon}.  The cross sections must additionally be multiplied by the appropriate leptonic branching ratio for the decay $E\to l^+l^-$.  These factors have been included when calculating the event rates, which assume a luminosity $\mathcal{L} = 2\times 10^{32}\,{\rm cm}^{-2}\,{\rm s}^{-1}$ at the Tevatron and $\mathcal{L} = 10^{34}\,{\rm cm}^{-2}\,{\rm s}^{-1}$ at the LHC.  No gap survival factor has been applied to these predictions.}
  \label{tab:results}
\end{table}

\begin{figure}
  \centering
  \includegraphics[width=0.89\textwidth]{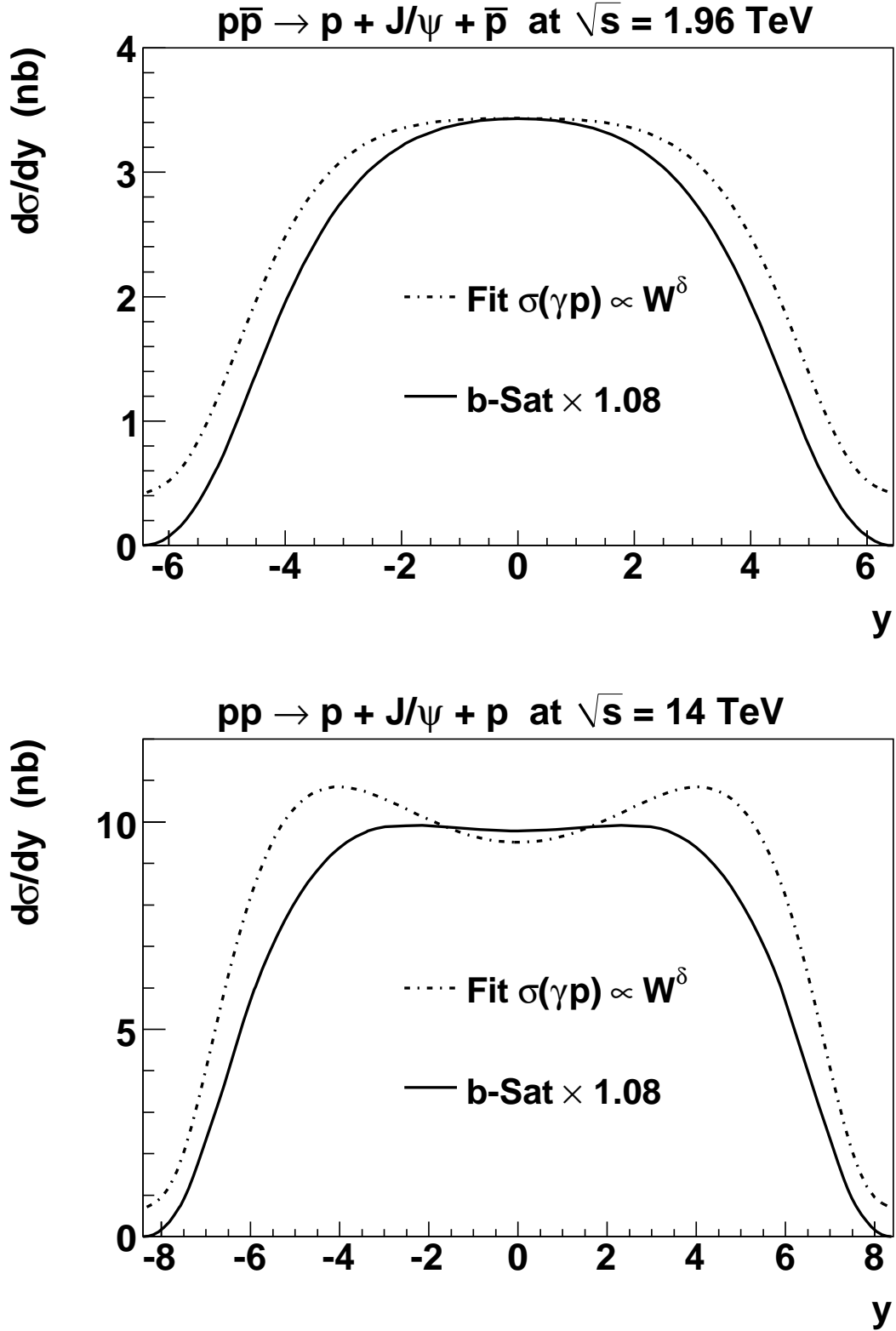}
  \caption{Rapidity distributions for exclusive photoproduction of $J/\psi$ mesons at the Tevatron and LHC.  The ``b-Sat'' model predictions are rescaled by a factor 1.08 to give optimum agreement with the HERA data \cite{Chekanov:2002xi,Aktas:2005xu}.  Also shown is the result of a direct fit to the HERA data of the form $\sigma(\gamma p) \propto W^\delta$.  No gap survival factor has been applied to these predictions.}
  \label{fig:dsdy_jpsi}
\end{figure}

\begin{figure}
  \centering
  \includegraphics[width=0.89\textwidth]{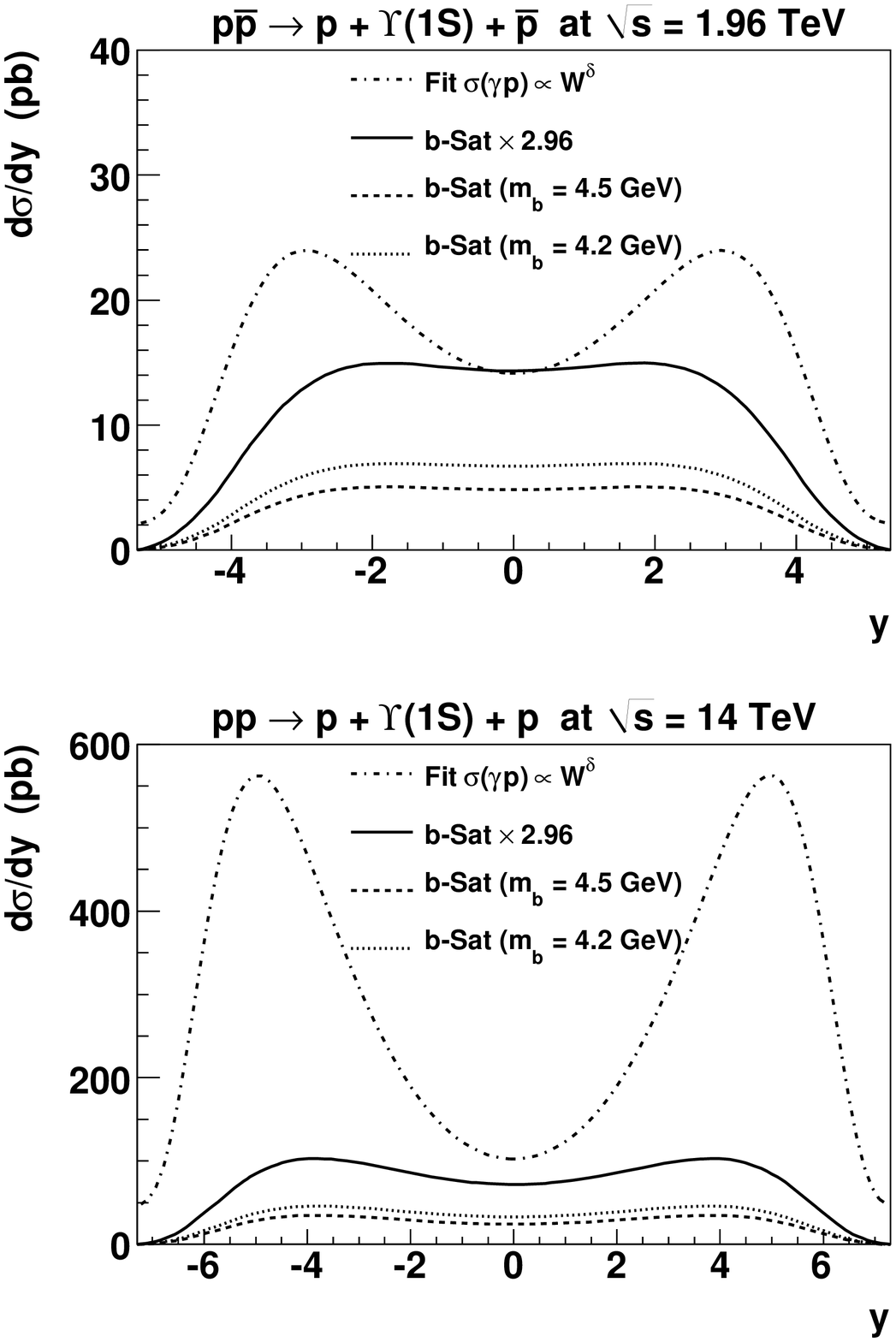}
  \caption{Rapidity distributions for exclusive photoproduction of $\Upsilon$ mesons at the Tevatron and LHC.  The ``b-Sat'' model predictions with $m_b = 4.5$ GeV are rescaled by a factor 2.96 to give optimum agreement with the HERA data \cite{Breitweg:1998ki,Adloff:2000vm,ZEUSupsilon}.  Also shown is the result of a direct fit to the HERA data of the form $\sigma(\gamma p) \propto W^\delta$.  No gap survival factor has been applied to these predictions.}
  \label{fig:dsdy_upsilon}
\end{figure}

\begin{figure}
  \centering
  \includegraphics[width=0.89\textwidth]{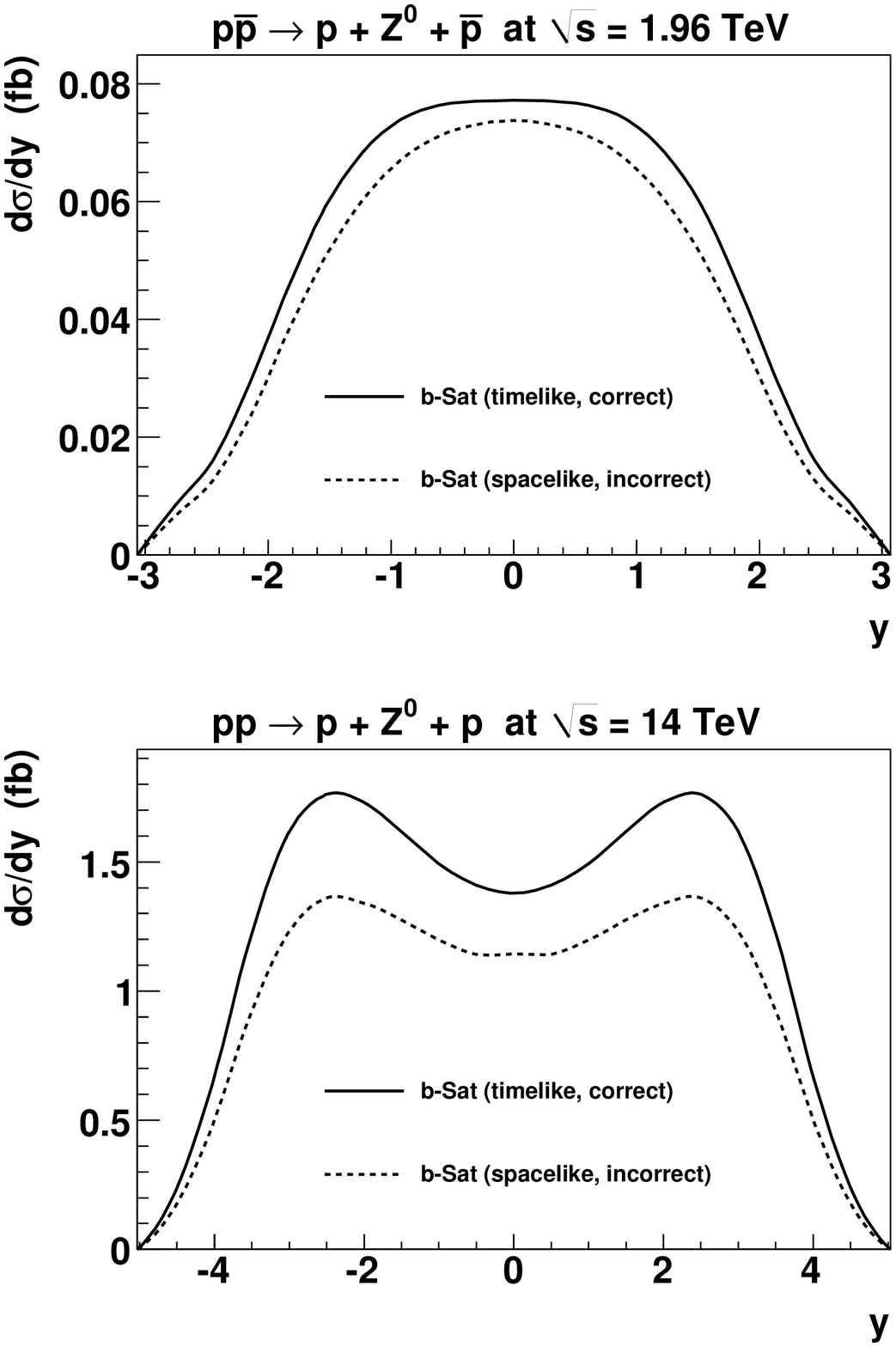}
  \caption{Rapidity distributions for exclusive photoproduction of $Z^0$ bosons at the Tevatron and LHC in the (correct) timelike case and the (incorrect) spacelike case.  No gap survival factor has been applied to these predictions.}
  \label{fig:dsdy_z}
\end{figure}

In Figs.~\ref{fig:dsdy_jpsi}, \ref{fig:dsdy_upsilon} and \ref{fig:dsdy_z} we show the rapidity distributions calculated using \eqref{eq:rapdist}.  Note the sensitivity of the $\Upsilon$ rapidity distribution at the LHC in Fig.~\ref{fig:dsdy_upsilon} to the $W$ dependence of the $\gamma p$ cross section: both the $W^\delta$ parameterisation and the rescaled b-Sat predictions describe the existing HERA data well, but give very different rapidity distributions.  Indeed, measurements of this distribution will provide an important constraint on the generalised gluon density \cite{Khoze:2008cx}.

For a given rapidity $y$, there are contributions from proton momentum fraction $x=(M_E/\sqrt{s})\exp(\pm y)$.  At central rapidity, the $x$ values probed at the LHC (Tevatron) are $2\times10^{-4}$ ($2\times10^{-3}$) for $J/\psi$, $7\times10^{-4}$ ($5\times10^{-3}$) for $\Upsilon$, and $7\times10^{-3}$ ($5\times10^{-2}$) for $Z^0$.  Therefore, apart from $Z^0$ production at the Tevatron, the $x$ values sampled for central rapidity are safely in the region $10^{-4}\lesssim x \lesssim 10^{-2}$ where the dipole cross section \eqref{eq:dsigmad2b} was fitted to HERA $F_2$ data.  However, moving further away from central rapidity, the dipole cross section must not only incorporate the correct dynamics at very small $x$, but also give sensible behaviour as $x\to 1$.  As noted in Ref.~\cite{Klein:2003vd}, the simple parameterisation $\sigma(\gamma p)\propto W^\delta$ gives rise to a discontinuity at threshold; see the dot-dashed curves in Figs.~\ref{fig:dsdy_jpsi} and \ref{fig:dsdy_upsilon}.  The b-Sat model is well behaved for large $x$, due to the damping factor $(1-x)^{5.6}$ in the input gluon distribution \eqref{eq:inputgluon}.  Note, however, that this form of the input leads to a very strong $x$ variation of the gluon density as $x\to 1$.  Obviously, this variation is not related to the evolution of the gluon density in the small-$x$ limit.  Therefore, in calculating the skewing and real part corrections, we divide by a factor $(1-x)^{5.6}$ when calculating the value of $\lambda$ in \eqref{eq:beta} and \eqref{eq:Rg} to approximately cancel the corresponding factor in the input gluon distribution \eqref{eq:inputgluon}; we set $\lambda=0$ if the value obtained lies outside the interval between 0 and 1.

\section{Discussion} \label{sec:discussion}

A similar approach to that used in this paper, that is, using the equivalent-photon approximation with the photon--proton cross sections given by the dipole model, has been applied in Ref.~\cite{Goncalves:2007sa} to calculate exclusive $J/\psi$ and $\Upsilon$ photoproduction at the LHC.  In that case, the impact parameter dependent version \cite{Kowalski:2006hc} of the colour glass condensate (CGC) model \cite{Iancu:2003ge} was used for the dipole cross section.  This ``b-CGC'' model has now been updated and discussed further \cite{Watt:2007nr}.  It is less successful than the ``b-Sat'' model in describing observables sensitive to relatively small dipole sizes, such as $F_2$ at high $Q^2$ and $J/\psi$ photoproduction at HERA, hence we do not use it in this paper.  The b-CGC model gives a less steep $W$ dependence than the b-Sat model, particularly for $\Upsilon$, and we find that we cannot reproduce the results of Ref.~\cite{Goncalves:2007sa}.\footnote{The results of Ref.~\cite{Goncalves:2007sa} for $\Upsilon$ are also inconsistent with the dipole model calculations compared to HERA $\Upsilon$ data in Ref.~\cite{Malka:2008}.}  Recently, an analysis of exclusive $\Upsilon$ photoproduction at hadron colliders has been performed within the $k_\perp$-factorisation framework in momentum space \cite{Rybarska:2008pk}.  We find good agreement between our results and those of Ref.~\cite{Rybarska:2008pk} for the $\gamma p\to \Upsilon+p$ cross section under similar model assumptions.

Early calculations of exclusive $Z^0$ photoproduction were made in Refs.~\cite{Bartels:1981jh,Pumplin:1996pi}.  More recently, a paper has appeared \cite{Goncalves:2007vi} taking a similar approach as in the present paper.  The total cross sections in Ref.~\cite{Goncalves:2007vi} are about a factor 5 larger than those presented here: such a large difference seems rather hard to explain by different model assumptions.\footnote{In fact, an erroneous additional factor of $2\pi$ in Ref.~\cite{Goncalves:2007vi} has since been discovered \cite{Machado:2008}.}  In that paper, another variant of the CGC model \cite{Iancu:2003ge} with a $t$-dependent saturation scale \cite{Marquet:2007qa} was used.  This model (and also the b-CGC model) are not well behaved for large $x$: the scattering amplitude does not vanish for $x \to 1$, and also the effective anomalous dimension $\gamma_{\rm eff}=\partial\ln\sigma_{q\bar{q}}/\partial\ln r^2$ blows up in this limit.  Hence it was necessary in Ref.~\cite{Goncalves:2007vi} to introduce a phenomenological correction factor, $(1-\bar{x})^5$, where $\bar{x}=(M_Z+m_p)^2/W^2$.  Note that Ref.~\cite{Goncalves:2007vi} did not include the correct timelike kinematics, but rather used the incorrect spacelike formula.  However, the numerical impact of this mistake is only about 20\% at the LHC, as shown in Fig.~\ref{fig:dsdy_z}.

Exclusive production of $J/\psi$ and $\Upsilon$ mesons via photon--Pomeron fusion was calculated within the $k_\perp$-factorisation framework in Ref.~\cite{Bzdak:2007cz} as a background to odderon--Pomeron fusion.  However, the main goal of Ref.~\cite{Bzdak:2007cz} was to provide an estimate of the odderon contribution to exclusive diffractive vector meson hadroproduction, and not to provide precise predictions for the photon-mediated process.  The estimates of the photon contribution were made there in order to monitor the impact of model assumptions in a relatively well-known situation.  The ``odderon to photon'' ratio for exclusive $Z^0$ production is expected to be strongly suppressed relative to the cases of exclusive $J/\psi$ or $\Upsilon$ production due to the smaller value of the strong coupling $\alpha_S(M_Z^2)$ compared to $\alpha_S(m_c^2)$ or $\alpha_S(m_b^2)$, and also due to a stronger Sudakov suppression for glueproduced $Z^0$ bosons because of the larger hard scale.

It should be noted that absorptive corrections are expected to dramatically alter the $t_i$ distributions, where $t_i\equiv (P_i-P_i^\prime)^2$, although this effect is washed out to a large extent in the $t_i$-integrated cross sections.  Indeed, it was proposed in Ref.~\cite{Khoze:2002dc} that the measurement of exclusive photon-exchange processes would provide a detailed probe of the rapidity gap survival probability.  It was estimated in Refs.~\cite{Schafer:2007mm,Rybarska:2008pk} that absorptive corrections reduce the $t_i$-integrated rapidity distributions by a factor $\sim$ 0.7--0.9.  Unfortunately, the inclusion of absorptive corrections requires knowledge of the transverse momenta of the outgoing protons, or equivalently the photon virtuality.  However, this is already integrated over in the photon flux of \eqref{eq:photonflux}.  Therefore, it is not possible to include absorptive corrections, or to present more detailed distributions involving the momenta of the final state particles, while maintaining the simple equivalent-photon approximation adopted here.  Instead, the rapidity distributions presented in this paper provide a reference point in the absence of absorptive corrections, and we refer to Refs.~\cite{Khoze:2002dc,Schafer:2007mm,Rybarska:2008pk,Khoze:2008cx} for a discussion of these effects, and for more differential distributions in the case of $J/\psi$ (and $\Upsilon$) production.  In principle, the methods of these papers could be applied using the b-Sat dipole model for the amplitude $\mathcal{A}(\gamma p\to E+p)$, including proper treatment of the photon polarisation, rather than relying on an extrapolation of HERA data.

\section{Conclusions} \label{sec:conclusions}

We have made predictions for the rapidity distributions of exclusive photoproduced $J/\psi$ and $\Upsilon$ mesons, and $Z^0$ bosons, expected at the Tevatron and LHC.  We used the equivalent-photon approximation with the photon--proton cross sections given by the impact parameter dependent dipole saturation model \cite{Kowalski:2003hm,Kowalski:2006hc}.  The normalisation of the $J/\psi$ and $\Upsilon$ predictions has been adjusted to give the best description of the available HERA data.  For the $Z^0$ case we cannot rely on existing data; however, the b-Sat model already describes well the similar process of DVCS at HERA \cite{Kowalski:2006hc,Watt:2007nr}.  We pointed out a crucial difference between the amplitude for DVCS and the amplitude for exclusive $Z^0$ photoproduction, and we have derived the wave functions for the case of a timelike $Z^0$ boson or virtual photon.  We have given a simple parameterisation of the photon--proton cross sections for use in future studies, and we have discussed the uncertainties inherent in our predictions.

Work is in progress on measurements of exclusive $J/\psi$ and $\Upsilon$ production at the Tevatron \cite{Pinfold:2008}.  Exclusive photoproduction processes will be important at the LHC, particularly if proton taggers are installed in the 420 m region \cite{Albrow:2005ig}.  Exclusive production of $\Upsilon$ mesons should be measured in the early days of LHC running, extending and improving the existing data from HERA, and providing valuable constraints on the shape of the dipole cross section or generalised gluon density \cite{Khoze:2008cx,Ovyn:2008}.  It is very unlikely that exclusive $Z^0$ photoproduction will be observed at the Tevatron, and it remains to be seen whether this process is observable at the LHC given the low event rate.

\begin{acknowledgments}
We thank Mike Albrow for suggesting that we perform the calculations in this paper.  We are grateful to Henri Kowalski and Robert Thorne for valuable discussions.  G.W.~acknowledges the UK Science and Technology Facilities Council for the award of a Responsive Research Associate position, and L.M.\ gratefully acknowledges the support of the DFG grant SFB~676.
\end{acknowledgments}

\appendix*
\section{Derivation of $Z^0/\gamma^*$ wave functions}

The dipole model approach has so far been used to describe processes with a photon or heavy boson having negative or vanishing virtuality, $q^2=-Q^2 \leq 0$.  In the case of $Z^0$ photoproduction, $\gamma p\to Z^0+p$, or timelike Compton scattering, $\gamma p\to \gamma^*+p$, it is needed to generalise the formula to the case of $q^2 = M^2>0$.

In the dipole model the light-cone wave function of the vector boson~$V$ is defined by elementary diagrams entering the transition amplitude $V(q)\,g(k) \to q(p_1)\,\bar q(p_2)$, where $q$, $k$, $p_1$ and $p_2$ denote the particle four-momenta.  The kinematics of this process at large energies in light-cone coordinates [for instance, $p = (p^+,p^-,\vec p)$, where $p^\pm = p^0 \pm p^z$] are defined as
\begin{equation}
  q = (q^+, M^2 / q^+, \vec 0), 
  \quad p_1 = \left(zq^+, \frac{{\vec p}_1 ^2 + m_f^2}{zq^+},\vec{p}_1\right), 
  \quad p_2 = \left((1-z)q^+, \frac{{\vec p}_2 ^2 + m_f^2}{(1-z)q^+},\vec{p}_2 \right),
\end{equation}
and $k=p_1+p_2-q$.  The three polarisation vectors of the boson are
\begin{equation}
  \epsilon_{\pm} = (0,0,\vec{\epsilon}_{\pm})\quad \mbox{where} \quad
  \vec{\epsilon}_{\pm} = \mp \frac{1}{\sqrt{2}} (1,\pm {\rm i}), 
\end{equation}
and 
\begin{equation}  
  \epsilon_{0} = \left(\frac{q^+}{M},-\frac{M}{q^+},\vec 0 \right),
\end{equation}
where $\epsilon_{\pm}$ and $\epsilon_{0}$ describe the transverse and longitudinal polarisations respectively. The calculation is performed in the high energy limit, that is, in the limit where $q^+$ is much larger than all other scales.

The vector boson wave function in momentum space may be obtained from a single Feynman diagram describing the $V(q)\,g(k)\to q_f(p_1)\,\bar q_f(p_2)$ subamplitude (see, for example, Refs.~\cite{Lepage:1980fj,Bartels:2007aa}) in which the gluon couples, for instance, to the quark line:
\begin{equation} \label{eq:wavemom}
  \tilde\Psi_{\lambda} ^{\lambda_1\lambda_2} (\vec{p},z) \; = \; 
	    {\rm i}{\cal N}\; \frac{zq^+\;\bar u_{\lambda_1}(p_1)\,\hat{\cal O}_\lambda\, v_{\lambda_2}(p_2) 
	    }{ (q-p_2)^2 - m_f^2 + {\rm i}\epsilon},  
\end{equation}
where $\,\vec{p}_1 = -\vec{p}_2 = \vec{p}\,$, $\hat {\cal O}_\lambda$ is a Dirac matrix characterising the coupling of the quark line to the incoming particle, and the spinors are taken in the helicity basis.  The normalisation factor ${\cal N}$ depends upon the convention: we will fix it later on so that the wave functions obtained match the standard dipole model expressions for the virtual vector boson with $q^2 \leq 0$.  The factor $zq^+$ in the numerator comes from the eikonal coupling of the gluon to the quark.\footnote{In the case of the gluon coupled to the antiquark, the numerator of \eqref{eq:wavemom} changes into $-(1-z)q^+$, and the denominator into $(q-p_1)^2 - m_f^2 + {\rm i}\epsilon$, but the final result remains the same up to a minus sign that comes from the scattering part and not the wave function part.} 

Let us first analyse the case of the timelike virtual photon production with $q^2 = M^2 > 0$.  For the photon, the operator $\hat {\cal O}_\lambda^{\gamma}$ takes the following form:
\begin{equation}\hat {\cal O}_\lambda^{\gamma} =  
  -{\rm i}e e_f \epsilon_{\lambda} ^{\mu} \gamma_{\mu}.
\end{equation}
In the high energy limit, the matrix element $\bar u_{\lambda_1}(p_1) \, \hat{\cal O}_\lambda^{\gamma}\, v_{\lambda_2}(p_2)$ for transverse polarisations, $\epsilon_{\pm}$, does not depend on the virtuality of the incoming photon, so this part of the wave function is the same for the spacelike and timelike virtual photons.  The virtual quark propagator, however, is sensitive to the photon virtuality.  In the timelike case it takes the form
\begin{equation}
  (q-p_2)^2 - m_f^2 + {\rm i}\epsilon \; = \; 
  \frac{1}{1-z}
  \left[-\vec{p}^2 - m_f^2  + z(1-z)M^2 +  {\rm i}\epsilon\right],
\end{equation}
to be compared with the standard expression for the photon with $q^2 = -Q^2<0$:
\begin{equation}
  (q-p_2)^2 - m_f^2 + {\rm i}\epsilon \; = \; 
  \frac{1}{1-z}\left[
    - \vec{p}^2 - m_f^2 - z(1-z)Q^2 +  {\rm i}\epsilon \right].
\end{equation}
In the spacelike photon case one defines a variable $\varepsilon^2 = Q^2z(1-z) + m_f^2>0$.  An analogous variable in the timelike photon case reads $\tilde\varepsilon^2 = m_f^2 - M^2 z(1-z) + {\rm i}\epsilon$, and its sign is not positive definite; for a large $M/m_f$ ratio $\tilde\varepsilon^2$ will be negative, except in the vicinity of the $z$~end-points.  
 
The wave function in coordinate space is obtained by taking the Fourier transform:
\begin{equation}
  \Psi_\lambda^{\lambda_1\lambda_2}(\vec{r},\ldots) = \int\!\frac{\dif^2\vec{p}}{(2\pi)^2}\;\exp({\rm i}\vec{p}\cdot\vec{r})\,\tilde\Psi_\lambda^{\lambda_1\lambda_2}(\vec{p},\ldots).
\end{equation}
Using the standard identities:
\begin{equation}
  \int_0^\infty\!\dif{p}\;p \, \frac{J_0(pr)}{p^2 + a^2} = K_0(ar) \;\;
  \mbox{for Re}\,a>0, 
\end{equation}
and
\begin{equation}
  \int\!\dif^2{\vec{p}}\;\vec{p} \, \tilde f(p) \, \exp({\rm i}\vec{p}\cdot\vec{r}) = 
  -{\rm i}\frac{\partial}{\partial \vec{r}}  
  \int\!\dif^2{\vec{p}} \; \tilde f(p) \, \exp({\rm i}\vec{p}\cdot\vec{r}),
\end{equation}
and imposing the normalisation convention used in Refs.~\cite{Kowalski:2003hm,Kowalski:2006hc} where a factor of $1/(4\pi)$ appears in the integration measure, see \eqref{eq:exclamp}, we find the expressions for the wave function of the transversely polarised virtual photon:
\begin{equation}
  \Psi^{\lambda_1\lambda_2} _{\lambda=\pm 1}(\vec{r},z,Q) =
  -e_f e \, \sqrt{2N_c}\,
  \left\{\pm
  \mathrm{i}e^{\pm \mathrm{i}\theta_r}[
    z\delta_{\lambda_1,\pm}\delta_{\lambda_2,\mp} - 
    (1-z)\delta_{\lambda_1,\mp}\delta_{\lambda_2,\pm}] \partial_r \,+ \, 
  m_f \delta_{\lambda_1,\pm}\delta_{\lambda_2,\pm}
  \right\}\, \frac{K_0(\varepsilon r)}{2\pi}
  \label{tspinphots}
\end{equation}
for the spacelike photon, and
\begin{equation}
  \Psi^{\lambda_1\lambda_2} _{\lambda=\pm 1}(\vec{r},z,M) =
  -e_f e \, \sqrt{2N_c}\,
  \left\{
  \pm \mathrm{i} e^{\pm \mathrm{i}\theta_r}[
    z\delta_{\lambda_1,\pm}\delta_{\lambda_2,\mp} - 
    (1-z)\delta_{\lambda_1,\mp}\delta_{\lambda_2,\pm}] \partial_r \,+ \, 
  m_f \delta_{\lambda_1,\pm}\delta_{\lambda_2,\pm}
  \right\}\, \frac{K_0(\tilde\varepsilon r)}{2\pi}
  \label{tspinphotl}
\end{equation}
for the timelike photon, where
\begin{equation}
\tilde\varepsilon = \begin{cases}\sqrt{m_f^2 - M^2 z(1-z)} & :\quad
m_f^2 - M^2 z(1-z) > 0\\
-{\rm i}\sqrt{M^2 z(1-z)-m_f^2} & : \quad
M^2 z(1-z) -m_f^2 > 0
\end{cases},
\end{equation}
and where $\partial_r K_0(\varepsilon r) = -\varepsilon K_1(\varepsilon r)$.  Note that the signs of the various terms of \eqref{tspinphots} differ from the wave functions given in Refs.~\cite{Kowalski:2003hm,Kowalski:2006hc}, but the result for the overlap function summed over all helicities is the same.

In the case of the longitudinal polarisation of the virtual photon the quark propagator is the same as above. The vertex part, however, is now sensitive to the incoming photon virtuality $q^2$.  One obtains 
\begin{align}
  \frac{z\, \bar u_{\lambda_1}(p_1) \, \varepsilon_{0} ^{\mu} \gamma_{\mu}
    \, v_{\lambda_2}(p_2) 
  }{
    (q-p_2)^2 - m_f^2 + {\rm i}\epsilon }
  & = 
  \frac{\sqrt{z(1-z)}}{Q} 
  \; \frac{\vec{p}^2 + m_f^2 - z(1-z) Q^2 }{ \vec{p}^2 + m_f^2 + z(1-z) Q^2 }
  \;\delta_{\lambda_1,-\lambda_2} \notag \\
  & =\; \frac{\sqrt{z(1-z)} }{Q}\; 
  \left[ 1 - \frac{2z(1-z) Q^2 }{ \vec{p}^2 + m_f^2 + z(1-z) Q^2 }\right]
  \;\delta_{\lambda_1,-\lambda_2}
\end{align}
for the spacelike photon, and
\begin{align}
  \frac{z\, \bar u_{\lambda_1}(p_1) \, \varepsilon_{0} ^{\mu} \gamma_{\mu}
    \, v_{\lambda_2}(p_2) 
  }{
    (q-p_2)^2 - m_f^2 + {\rm i}\epsilon } & = 
  \frac{\sqrt{z(1-z)} }{ M} 
  \; \frac{\vec{p}^2 + m_f^2 + z(1-z) M^2 }{ \vec{p}^2 + m_f^2 - z(1-z) M^2 - {\rm i}\epsilon } 
  \;\delta_{\lambda_1,-\lambda_2} \notag \\
  & =\; \frac{\sqrt{z(1-z)} }{ M}\; 
  \left[ 1 + \frac{2z(1-z) M^2 }{ \vec{p}^2 + m_f^2 - z(1-z) M^2 - {\rm i}\epsilon }
    \right] 
  \;\delta_{\lambda_1,-\lambda_2}
\end{align}
for the timelike photon.  As usual, the factors $\sqrt{z(1-z)}$ appearing in the wave functions of the initial and final state are absorbed into the phase space integrations in the impact factor. The constant terms (equal to 1) in the square brackets cancel in the calculation of the impact factor due to gauge invariance (the contributions of the quark and antiquark scattering enter the impact factor calculation with the opposite phase).  It is now straightforward to obtain the final results for the longitudinally polarised photon:
\begin{equation}
  \Psi^{\lambda_1\lambda_2} _{\lambda=0}(\vec{r},z,Q) =  e_f e \, \sqrt{N_c}\, 
  \delta_{\lambda_1,-\lambda_2} \, 2Qz(1-z)\, 
  \frac{K_0(\varepsilon r)}{2\pi}
  \label{lspinphots}
\end{equation}
for the spacelike photon, and
\begin{equation}
  \Psi^{\lambda_1\lambda_2} _{\lambda=0}(\vec{r},z,M) =  -e_f e \, \sqrt{N_c}\, 
  \delta_{\lambda_1,-\lambda_2} \, 2Mz(1-z)\, 
  \frac{K_0(\tilde\varepsilon r)}{2\pi}
  \label{lspinphott}
\end{equation}
for the timelike photon.

In the case of the $Z^0$ boson, the coupling to quarks contains both vector and axial-vector parts.  The amplitude for the $Z^0 \to q_f\bar q_f$ transition is described by
\begin{equation}
  T(Z^0(\lambda)\to q_f\bar q_f) = 
  -\frac{{\rm i}e }{ \sin 2\theta_W} \,
  \varepsilon_\lambda ^\mu \,
  \bar u_f\, [\, g^f _v \gamma_{\mu} - g^f _a \gamma_{\mu} \gamma_5 \,]\, v_f, 
\end{equation}
thus
\begin{equation}
  \hat{\cal O}_\lambda^{Z^0} = -\frac{{\rm i}e }{ \sin 2\theta_W} \,
  \varepsilon_\lambda ^\mu \,[\, g^f _v \gamma_{\mu} - g^f _a \gamma_{\mu} 
    \gamma_5 \,],
\end{equation}
where $u_f$ and $v_f$ are the spinors of the quark and antiquark of the flavour $f$, and $\theta_W$ is the Weinberg angle.  The vector couplings are
\begin{equation}
  g^{u,c,t} _v = \frac{1}{2} - \frac{4}{3} \sin^2 \theta_W \qquad\mbox{and}\qquad
  g^{d,s,b} _v = -\frac{1}{2} + \frac{2}{3} \sin^2 \theta_W,
\end{equation}
while the axial-vector couplings are
\begin{equation}
  g^{u,c,t} _a = \frac{1}{2} \qquad\mbox{and}\qquad
  g^{d,s,b} _a = -\frac{1}{2}.
\end{equation}
It is natural to decompose the $Z^0$ wave function into distinct vector and axial-vector parts:
\begin{equation}
  \Psi_{\lambda} ^{\lambda_1 \lambda_2}(\vec{r},z,M) = 
  V_{\lambda} ^{\lambda_1 \lambda_2}(\vec{r},z,M) 
  - A_{\lambda} ^{\lambda_1 \lambda_2}(\vec{r},z,M).
\end{equation}
The analysis of the virtual photon case gives immediately the vector part of the timelike $Z^0$ wave function:
\begin{equation}
  V^{\lambda_1 \lambda_2} _{\pm 1}(\vec{r},z,M) = 
  \,-\frac{e g^f _v }{ \sin 2\theta_W}\, \sqrt{2N_c}\,
  \left\{\pm
  \mathrm{i}e^{\pm \mathrm{i}\theta_r}[
    z\delta_{\lambda_1,\pm}\delta_{\lambda_2,\mp} - 
    (1-z)\delta_{\lambda_1,\mp}\delta_{\lambda_2,\pm}] \partial_r \, + \,   m_f \delta_{\lambda_1,\pm}\delta_{\lambda_2,\pm}
  \right\}\, \frac{K_0(\tilde\varepsilon r)}{2\pi},
  \label{tspinz0vt}
\end{equation}
\begin{equation}
  V^{\lambda_1\lambda_2} _{0}(\vec r,z,M) =  
  \,-\frac{e g^f _v }{ \sin 2\theta_W}\, 
  \frac{\sqrt{N_c} }{ M}\, 
  \delta_{\lambda_1,-\lambda_2} \, 2M^2 z(1-z)\,
  \frac{K_0(\tilde\varepsilon r)}{2\pi}. 
  \label{lspinz0vs}
\end{equation}
The axial-vector part may be obtained in a similar way as the vector part; one needs to take into account  that $\gamma_5 u_{\lambda} = 2\lambda v_{-\lambda}$ (where $\lambda = \pm 1/2$).  Thus, one gets
\begin{equation}
  A^{\lambda_1 \lambda_2} _{\pm 1}(\vec{r},z,M) = 
  \, \frac{e  g^f _a }{ \sin 2\theta_W}\, \sqrt{2N_c}\,
  \left\{
  -\mathrm{i}e^{\pm\mathrm{i}\theta_r}[
    z\delta_{\lambda_1,\pm}\delta_{\lambda_2,\mp} + 
    (1-z)\delta_{\lambda_1,\mp}\delta_{\lambda_2,\pm}] \partial_r \, 
  \pm \, m_f (1-2z) \delta_{\lambda_1,\pm}\delta_{\lambda_2,\pm}
  \right\}\, \frac{K_0(\tilde\varepsilon r)}{2\pi},
  \label{tspinz0at}
\end{equation}
\begin{equation}
  A^{\lambda_1\lambda_2} _{0}(\vec r,z,M) =  
  \,-\frac{e g^f _a }{ \sin 2\theta_W}\, \frac{\sqrt{N_c}}{ M}\, 
  \left\{ \,
  \delta_{\lambda_1,-\lambda_2} \, 2\lambda_1\;
  \left[\,2M^2 z(1-z) \, + 2m_f ^2\, \right] - \mathrm{i} \delta_{\lambda_1,\lambda_2} e^{-\mathrm{i}2\lambda_1\theta_r}
  2m_f\partial_r \;\right\}
  \frac{K_0(\tilde\varepsilon r)}{2\pi}. 
  \label{lspinz0as}
\end{equation}
The analogous formulae for the spacelike $Z^0$ is obtained by the replacements $M \to -Q$ and $\tilde\varepsilon \to \varepsilon$, then one recovers the formulae of Fiore and Zoller \cite{Fiore:2005yi,Fiore:2005bp} up to an overall factor of $1/\sqrt{4\pi}$ which in our case is included in the integration measure rather than the wave function; see \eqref{eq:exclamp}.

The contribution to $Z^0$ photoproduction in the forward direction comes only from the vector part of the $Z^0$ wave function.  This follows from the fact that the vector and axial-vector currents have opposite parity.  Therefore, the overlap function for $Z^0$ photoproduction $(\Psi^*_{Z^0}\Psi_{\gamma})_T^f $ follows from the expression for timelike Compton scattering $(\Psi^*_{\gamma^*(M_Z^2)} \Psi_{\gamma})_T^f$ after the appropriate coupling adjustment, $e e_f \to \frac{e g_v ^f }{ \sin 2\theta_W}$.

\end{document}